\documentstyle[epsf,prb,twocolumn,aps]{revtex}

\begin{document}

\draft

\wideabs{

\title{Spin- and charge-density oscillations in spin chains and quantum wires}

\author{Stefan Rommer}

\address{Department of Physics and Astronomy, University of California,
Irvine, California 92697}

\author{Sebastian Eggert}

\address{Institute of Theoretical Physics, Chalmers University of
Technology and G\"oteborg University, 41296 G\"{o}teborg, Sweden}
\date{Submitted: December 7, 1999.  Last change: \today}

\maketitle

\begin{abstract}
We analyze the spin- and charge-density oscillations near 
impurities in spin chains and quantum wires.  These so-called
Friedel oscillations give detailed information about the 
impurity and also about the interactions in the system.
The temperature dependence of these oscillations 
explicitly shows the renormalization of backscattering and
conductivity, which we analyze for a number of different 
impurity models. We are also able to analyze screening effects
in one dimension.  The relation to the Kondo effect 
and experimental consequences are discussed. 
\end{abstract}

\pacs{PACS numbers: 71.10.Pm, 75.10.Jm, 72.15.Qm, 73.23.-b}



}\narrowtext
\section{Introduction} \label{introduction}
There is growing interest in impurities in low-dimensional electron 
and magnetic systems spurred by high temperature superconductivity
and experimental progress in producing ever smaller electronic
structures.  There appears to be two central aspects that are studied
most in this context, namely the effect of impurities on the transport
properties in mesoscopic systems on the one hand,\cite{kane1} and
impurity-impurity interactions in antiferromagnetic systems due to
impurity induced magnetic order\cite{elbio} on the other hand.
In this paper we show that the charge- and spin-densities near impurities
give a great deal of information about both of those aspects and
allow us to study a number of impurity models in one dimension 
in detail.

Induced density fluctuations at twice the Fermi wave-vector, 
so-called Friedel oscillations,\cite{friedel} are
a common impurity effect in fermionic systems, which
are enhanced in lower dimensions. 
  There are two distinct physical effects that
can give rise to Friedel oscillations.  The most common source 
is a simple {\it interference effect}
 as considered in the original work by Friedel.\cite{friedel}
Fermions scatter off the impurity, resulting in  
a superposition of incoming and outgoing wave-functions.
Summing up the squares of the corresponding wave-functions 
up to the sharp cutoff at the
Fermi wave-vector $k_F$ results in a characteristic interference
pattern with a $2 k_F x$ modulation, namely the Friedel oscillations.
Clearly, this pattern can give a great deal of information about the 
impurity, in particular details about the scattering process.
A second source for the $2 k_F x$ oscillations are {\it interaction
effects} due to the screening of an impurity with a net charge or a
magnetic moment.  A typical example of this effect is the
Kondo screening cloud,\cite{sorensen} which we also analyze in
this paper.  The $2 k_F x$ oscillations due to screening have typically
a different characteristic amplitude as a function of $x$ than 
those due to backscattering, as we will discuss in more detail below.

We now consider the density oscillations in one-dimensional systems
such as spin-chains and interacting quantum wires (Luttinger Liquids) 
in order to understand the detailed
effects of impurity scattering and screening as a function of temperature.
In the classic work by Kane and Fisher\cite{kane1,kanereview} it was found that
a generic impurity in a spinless Luttinger Liquid results in a
renormalization of the conductivity with temperature, which leads to
a perfectly reflecting barrier at $T=0$ for repulsive interactions.
Interestingly, this behavior can also be explained in terms of repeated
scattering off the Friedel oscillations,
which gives an explicit expression of the transmission coefficient 
in the weak coupling limit.\cite{yue}
Independently, the analogous renormalization behavior was also found in the 
spin-1/2 chain,\cite{eggert} where a generic 
perturbation in the chain effectively renormalizes to an open boundary 
condition as $T\to 0$.  
However, it is possible that a special symmetry in the Hamiltonian
reverses this renormalization, which leads to resonant tunneling in quantum
wires\cite{kanereview,kane2} or the healing of a two-link problem in the 
spin-1/2 chain.\cite{eggert}  The renormalization behavior in 
that case is analogous to the two-channel Kondo effect.\cite{eggert,2CK}  

The renormalization flow can easily be tested numerically 
by examining the scaling of the finite size energy gaps,\cite{eggert,qin}
but we now would like to determine the reflection coefficient 
directly by analyzing the induced density oscillations which are 
also interesting in their own right.  In addition, we also consider 
the density oscillations from impurity models near an edge, 
impurities with a net charge or magnetic moment (Kondo-type impurities), 
and integrable impurities.  The detailed renormalization of the 
impurity backscattering as well as screening can
be studied in each case by analyzing the induced density oscillations
as a function of temperature, which we determined numerically 
with the  Transfer Matrix 
Renormalization Group (TMRG) for impurities.\cite{2CK,rommer,TMDMRG}
This allows us to make predictions for conductivity
measurements in quantum wires and for Knight shift measurements
in spin chains, e.g.~Nuclear Magnetic Resonance (NMR) experiments.
In all cases we find a typical renormalization to a
fixed point of the Luttinger Liquid model, which is described
in terms of a simple (open or periodic) boundary condition in agreement 
with field theory calculations.

The rest of this paper is organized as follows.
In Sec.~\ref{model} we present the model Hamiltonian and review the
results for Friedel oscillations due to an open end (i.e.~complete
backscattering). 
Different impurity models of a modified link, two modified links,
an edge impurity, Kondo impurities, and an integrable impurity are 
then analyzed in detail in Sec.~\ref{impurities}. 
Section~\ref{numerics} contains a description of the
numerical methods used and a critical discussion about the possible
numerical errors.  
We conclude with a summary and a discussion about 
experimental relevance in Sec.~\ref{summary}.

\section{The Model} \label{model}
The standard model we are considering here are spinless interacting fermions
on a one-dimensional lattice, described by the Hamiltonian
\begin{equation}
H  =  \sum_{i} \left[-t (\Psi_i^\dagger \Psi_{i+1}^{} +
\Psi_{i+1}^\dagger \Psi_{i}^{}) + U n_i n_{i+1} - \mu n_i \right],
\label{fermions}
\end{equation}
where $n_i = \Psi_i^\dagger \Psi_{i}^{}$ is the fermion density.
Although this Hamiltonian neglects the spin degrees of freedom of real 
electrons in quantum wires, it captures the essential physics in
conductivity experiments.  Moreover, this model is equivalent to 
the spin-1/2 chain 
\begin{equation}
H =\sum_i \left[\frac{J}{2}(S_i^+ S_{i+1}^- + S_i^- S_{i+1}^+) 
+ J_z S_i^z S_{i+1}^z - B S_i^z\right]
\label{heisenberg}
\end{equation}
where the spin operators are related to the fermion field by the
Jordan-Wigner transformation
\begin{equation}
S_i^z = n_i - \case{1}{2}, \ \ \ S_i^- =  (-1)^i \Psi_i \exp{i \pi \sum_j^{i-1}n_j},
\label{JWtrafo}
\end{equation}
with $J = 2 t, \ J_z = U$ and $B= \mu - U$.

The model in Eq.~(\ref{fermions}) can be analyzed by standard bosonization
techniques in the low-temperature limit.  For low energies we only
consider excitations around the Fermi-points $\pm k_F$ and introduce
left- and right-moving fermion fields with a linear dispersion relation
\begin{equation}
\Psi(x) = 
e^{-ik_Fx}\psi_{L}(x) + e^{ik_Fx}\psi_{R}(x). \label{linearization}
\end{equation}
The chiral fermion fields can then be bosonized using the usual 
bosonization rules
\begin{equation}
\psi_{L/R}^\dagger \psi_{L/R}^{} = \case{1}{\sqrt{4 \pi}} \left(\partial_x \phi
\pm \Pi_\phi\right),
\end{equation}
where $\Pi_\phi$ is the conjugate momenta to the boson field $\phi$.
This  results in the following boson Hamiltonian density
\begin{equation}
{\cal H}  \ =  \     \case{v}{2}
\left[g^{-1}({\partial_x \phi})^2 +  g \Pi_\phi^2\right],\label{boson}
\end{equation}
which can be solved by a simple rescaling of the boson with the
interaction parameter $g$.   The parameter $g$ and the
velocity $v$ can in principle be calculated for any  interaction
strength $U$ and chemical potential $\mu$ with Bethe ansatz 
techniques.\cite{korepin}  To lowest order in $U$ we get 
$g = 1- 2 U/\pi v$ and $v = \sqrt{4 t^2-\mu^2} + 2 U/\pi$,
so that  $g<1$ for repulsive interactions.

We now want to analyze the density oscillations using this formalism.
Already from the decomposition of the fermion field in 
Eq.~(\ref{linearization}) it is clear that the fermion density
may contain an oscillating component with $2 k_F x$.  To see this
explicitly we can
write the charge density in quantum wires (or equivalently the 
spin density $\langle S_z \rangle$ in spin chains)
in terms of left- and right-movers 
\begin{eqnarray}
\langle \Psi^{\dagger} \Psi^{}\rangle & = &
\langle \psi_{L}^{\dagger}\psi_{L}^{} \rangle +
\langle \psi_{R}^{\dagger}\psi_{R}^{} \rangle  \nonumber \\
& & \ \ \ +
e^{i 2k_F x} \langle \psi_{L}^{\dagger}\psi_{R}^{} \rangle +
e^{-i 2k_F x} \langle \psi_{R}^{\dagger}\psi_{L}^{} \rangle.
\label{density}
\end{eqnarray}
The first two uniform terms just represent the overall fermion density
in the bulk system, while the last two ``Friedel'' terms are the 
density oscillations $n_{\rm osc}$ we are interested in.  
In a system with translational invariance
the left- and right-moving fields are uncorrelated 
$\langle \psi_{L}^{\dagger}\psi_{R}^{} \rangle =0$ and no 
density oscillations are present.  An impurity, however,  scatters 
left- into right-movers and the amplitude of the oscillations 
gives detailed information about the backscattering.

As the simplest example of this effect, let us consider an open boundary,
i.e.\ an impurity with complete backscattering at the origin.   
In this case the correlation functions can be calculated 
directly.\cite{eggert,mattsson,boundcorr,fabrizio,eggopen,finiteT}  
For the particular case of the left-right correlation function at 
equal space and time we find
\begin{equation}
\langle  \psi_{L}^{\dagger}(x)\psi_{R}^{}(x) \rangle  
\propto \left(\frac{\pi T}{v \sinh{2 \pi x T/v}}\right)^g,
\label{LR}
\end{equation}
so that the density oscillations are given by
\begin{equation}
n_{\rm osc} \propto \sin(2 k_F x)
\left(\frac{\pi T}{v \sinh{2 \pi x T/v}}\right)^g.
\label{n_osc}
\end{equation}
The Friedel oscillations are exponentially damped with 
temperature, because the  incoming and outgoing wave-functions that 
form the interference pattern lose coherence due to temperature fluctuations.
In the limit $T\to 0$ we recover the result of Ref.~\onlinecite{egger}
where a power-law decay of the Friedel 
oscillation $n_{\rm osc} \propto 1/x^{g}$ was predicted.

It is now important to realize that the fermions or spins are still
pinned to a lattice, i.e. $x = $ Integer,
which gives interesting additional effects.
In particular, at half-filling $k_F = \pi/2$ the Friedel oscillations
in Eq.~(\ref{n_osc}) are identically zero $\sin(\pi x) = 0$ 
for integer $x$, which can 
easily be understood from particle-hole symmetry (or equivalently
spin-flip symmetry).  Half-filling is a natural state for 
the spin chains in zero magnetic field, but a small magnetic field
changes the Fermi vector slightly $k_F = \pi/2 + B/v$.  In that case,
Eq.~(\ref{n_osc}) becomes
\begin{equation}
n_{\rm osc} \propto (-1)^x \sin(2 B x/v)
\left(\frac{\pi T}{v \sinh{2 \pi x T/v}}\right)^g.
\label{n_osc2}
\end{equation}
Now, the Friedel oscillations are simply alternating on the lattice
and for distances below the magnetic length scale $x < v/B$ we can use
$\sin(2 B x/v) \to 2 B x/v$ so that remarkably the oscillations actually
{\it increase} with $x^{1-g}$. This effect was first observed for
the Heisenberg chain ($J_z = J, \ \ g = 1/2, \ \ v=J \pi/2$), where the 
local susceptibilities $\chi(x)$ can be written as\cite{eggopen}
\begin{equation}
\chi(x) = \chi_0 \ - \ c\  (-1)^{x} \; \chi^{\rm bs}(x) , \label{susc}
\end{equation}
with the amplitude of the alternating part given by
\begin{equation}
\chi^{\rm bs}(x) = \frac{x \sqrt{T}}{\sqrt{\sinh 4xT}}.
\label{chi} 
\end{equation}
Here $\chi_0$ is the bulk susceptibility in the chain\cite{suscept}
and we measure $T$ in units of $J$.
The sign was chosen so that the (constant) overall amplitude $c$
of the alternating part is positive.
The superscript bs indicates that the alternating
susceptibility is due to backscattering.
As shown in Fig.~\ref{altsusc} from TMRG simulations
there is a characteristic maximum 
because the temperature damping eventually dominates over the increasing 
oscillations.  Clearly, the expression in Eq.~(\ref{chi}) reproduces 
the shape of this alternating part rather well,
although we have neglected possible logarithmic corrections 
(multiplicative and additive), which may be responsible for the 
apparent shift in the characteristic maximum in Fig.~\ref{altsusc}.  
The numerical TMRG results of the local susceptibility 
near the open end $\chi^{\rm bs}(x)$
will be used as the reference data for a completely backscattering 
impurity in our studies in the next section.  The numerical data 
automatically contains all corrections due to irrelevant higher order
operators.  The logarithmic corrections to Eq.~(\ref{chi})
due to the leading irrelevant 
operator have a special behavior near a boundary,\cite{brunel} which 
we have not tried to predict for the local susceptibility, 
but numerically we find that a possible 
multiplicative logarithmic correction for $\chi(x)$
appears to have a {\it negative} power of $\ln(x)$.
The maximum in Fig.~\ref{altsusc} occurs at $x \propto 1/T$ with an
amplitude $\chi_{\rm alt} \propto 1/\sqrt{T}$, which results in a 
characteristic feature in NMR experiments, so that it was possible to
confirm this effect experimentally as well.\cite{NMR}
At zero temperature the ground state has a staggered magnetization 
which has a maximum in the center of a finite chain (assuming
an odd number of sites).\cite{laukamp}
The magnetization for finite chains with impurities 
has also recently been analyzed, which resulted in interesting patterns
that reveal the nature of the strong correlations in the system.\cite{nishino}

Even for a partially reflecting impurity we expect that
the same alternating contribution as in Eq.~(\ref{chi}) due to backscattering
is present, but  with an amplitude $c$ that 
increases monotonically with the reflection coefficient $R$.
In fact we can make a firm connection between the relative amplitudes
and the reflection coefficients by considering free fermions $U=0$ for which 
we can find the eigenfunctions exactly even in the presence of impurities.
Clearly the eigenfunctions are given by plane wave solutions $|k\rangle$
which contain a special mix of left- and right-moving components due
to the impurity.  Just like without impurities there are in fact always two
such degenerate orthogonal solutions.  We found the solutions for 
generic impurity models and looked at the spatial structure of the
square of the wave-functions, which contains an interference pattern of
incoming and outgoing waves.  In general,  we always find 
\begin{equation}
\left| \langle x | k \rangle \right|^2 = \frac{1}{\pi} \left( 1 + 
\sqrt{R(k)} \cos (2 k x + 2 \Phi) \right), \label{wavefcn}
\end{equation}
where the summation over the two degenerate solutions is implied. Here
$R(k)$ is the ordinary k-dependent reflection coefficient which has been
determined 
independently according to text-book methods. Therefore, the magnitude
of the interference is exactly given by the square root of
the reflection coefficient, which is maybe not too surprising but very
useful in our analysis.  In particular, when we consider the fermion
density at half filling we are really directly looking 
at the spatial structure of the wave-function.  We can write
near half-filling (i.e. for a small field $B$
in the spin chain model) 
\begin{eqnarray}
n(x)-1/2 & = & \int_{\pi/2}^{\pi/2 + B/v} \left| \langle x | k \rangle
\right|^2 dk \nonumber \\ 
& \stackrel{B\to 0}{=} & \case{B}{v} 
\left| \langle x | \case{\pi}{2} \rangle \right|^2  
\nonumber \\
& = & B \left[ \chi_0 - c_R  
(-1)^x \chi^{\rm bs} \right],
\label{linresp}
\end{eqnarray}
where we have used the fact that the spin density for the Heisenberg chain
in a small field 
is just given by the susceptibility in Eq.~(\ref{susc}), but with
a coefficient $c_R$ which now depends on the reflection coefficient $R$
near half filling.  Together with Eq.~(\ref{wavefcn}) we therefore
arrive at the central result that at half-filling {\it the 
reflection coefficient is proportional to 
to the square of the alternating density amplitude}
\begin{equation}
R = \left( \frac{c_R}{c} \right)^2, \label{R}
\end{equation}
where $c=c_{R=1}$ is the coefficient corresponding to complete 
backscattering in Eq.~(\ref{susc}).
We use this formula to estimate the reflection coefficient from the density 
oscillations for various impurity models in the following.

{\begin{figure}
\epsfxsize=7cm
\centerline{\epsfbox{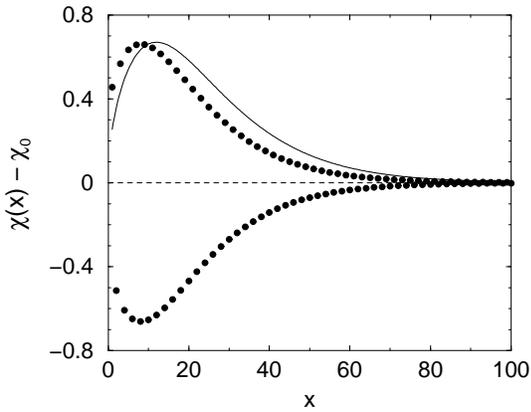}}
\caption{Local susceptibility close to the open end of a spin-1/2
chain from TMRG data for $T=0.04J$ compared to Eqs.~(\ref{susc}) 
and (\ref{chi}) 
with $c=0.51$, which was determined by matching the characteristic maxima.}
\label{altsusc}
\end{figure}}

As mentioned above there may also be $2 k_F x$ density oscillations due
to screening, so that the alternating susceptibility is in general a sum
of two parts
\begin{equation}
\chi^{\rm alt}(x)  \equiv   \chi(x) - \chi_0
 =   (-1)^x \left[\chi^{\rm screening}(x) -  c_R  \chi^{\rm bs}(x)\right].
\label{altchi}
\end{equation}
In the case of overscreening the 
neighboring spins (or electrons)  overcompensate the magnetic (or electric) 
impurity and leave an effective impurity with opposite moment 
which in turn gets screened by
the next nearest neighbors and so on.  This finally results in a
screening cloud. 
Screening is purely an interaction effect where a $2 k_F x$ density 
oscillation is induced by an ``active'' impurity Hamiltonian 
$\langle \psi_L^\dagger \psi_R^{} H_{\rm imp} \rangle \neq 0$.
The $2 k_F x$ oscillations due to backscattering, however, are purely an 
interference effect and are even present in non-interacting fermion systems.
The special shape and the increasing nature of the alternating part in 
Eq.~(\ref{chi}) for $g=1/2$
makes it possible to easily identify the contribution due
to backscattering, so that we can always separate the two possible effects
near half-filling.  In what follows we therefore always
use the special choice of coupling $U=2t$ 
corresponding to the Heisenberg model $J_z=J$. 
This model can be used to demonstrate the generic behavior of impurity 
effects in mesoscopic systems and also gives experimental consequences for 
spin-chain compounds.  The Luttinger Liquid
parameter takes the value $g=1/2$ in this case, which is the 
strongest possible interaction at half-filling before Umklapp
scattering becomes relevant.

\section{Impurity models} \label{impurities}

\subsection{One modified link} \label{onelink}

Maybe the simplest impurity to consider is a weak link in the chain,
i.e.\ a modified hopping $J'$ between two sites in the chain
as shown in Fig.~\ref{fig:oneweak}
\begin{equation}
H = -t \sum_{i\neq 0} (\Psi_i^\dagger \Psi_{i+1}^{} + 
\Psi_{i+1}^\dagger \Psi_{i}^{}) -
J' (\Psi_0^\dagger \Psi_{1}^{} +    
\Psi_{1}^\dagger \Psi_{0}^{}) .
\end{equation}

{\begin{figure}
\epsfxsize=7cm
\centerline{\epsfbox{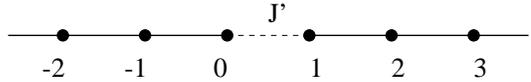}}
\caption{One modified link.}
\label{fig:oneweak}
\end{figure}}

The wave-functions and reflection coefficient $R(k)$ for this problem can 
be calculated exactly, with the result that 
\begin{equation}
R(k) =\frac{ t^4 -2 t^2 J'^2 + J'^4}{t^4 -2 t^2 J'^2 \cos 2 k + J'^4}.
\label{Rfree}
\end{equation}
However, once the interaction $U$ is introduced this problem becomes
highly non-trivial and the reflection coefficient renormalizes with
temperature $T$.  The interacting system has been studied in the context of 
both spinless fermions\cite{kane1} and the spin-1/2 chain,\cite{eggert}
where it was found that repulsive interactions $U > 0$ make the
perturbation of one link relevant, so that it renormalizes 
to a completely reflecting barrier as $T\to 0$. 
A small weakening of a link $J' \alt t$ 
produces a relevant backscattering operator in the periodic chain of 
scaling dimension $d=g$, so that this link effectively weakens further as
the temperature is lowered.  Below a cross-over temperature
$T_K$ (analogous to a Kondo-temperature) the link has weakened so 
much that it is more useful to consider the problem of two open ends
that are weakly coupled,  which is now described by an irrelevant operator
of scaling dimension $d=1/g$.  Therefore, this coupling weakens further
and ultimately the open boundary condition represents the stable fixed point
as $T\to 0$.  The same analysis is also true for a slight strengthening
of a link $J' \agt t$, because in this case the two ends lock into 
a ``singlet'' state as the effective coupling grows,
and the remaining ends are weakly coupled with a virtual
coupling of order $t^2/J'$ which is again irrelevant.

We consider the interacting system with $U = 2 t$, which we can write 
in terms of an SU(2) invariant spin Hamiltonian via the Jordan-Wigner
transformation in Eq.~(\ref{JWtrafo})  with a modified
Heisenberg coupling between two spins
\begin{equation}
H =  J \sum_{i\neq 0} {\bf S}_i \cdot {\bf S}_{i+1}
 + J^\prime {\bf S}_0 \cdot {\bf S}_1.
\label{oneweakham}
\end{equation}
We now want to analyze the density oscillation near the impurity
in order to extract the reflection coefficient as described above.
In Fig.~\ref{fig:oneweakaltpart} we show the amplitude of the 
alternating spin density for different coupling strengths $J'$.
Clearly the shape as a function of distance $x$ remains largely the 
same as in Fig.~\ref{altsusc} for all $J'$ so that the functional dependence
in Eq.~(\ref{chi}) is still adequate, but with an overall
coefficient $c$ which is now related to the reflection coefficient $R$
as postulated in Eq.~(\ref{R}).

{\begin{figure}
\epsfxsize=7cm
\centerline{\epsfbox{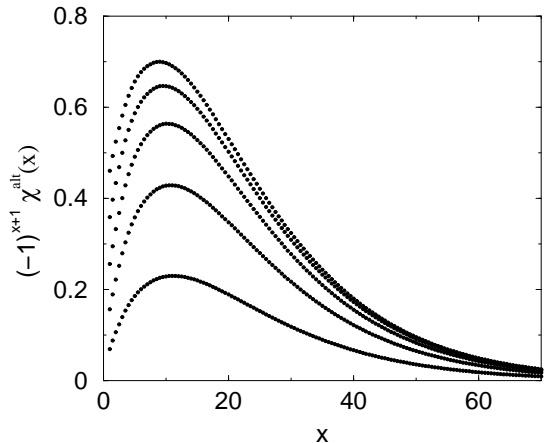}}
\caption{Envelope of the alternating susceptibility of
the one-link impurity at
$T=0.04J$ for  $J^\prime/J = 0.0, 0.2, 0.4,
0.6, 0.8$ from above. }
\label{fig:oneweakaltpart}
\end{figure}}

The reflection coefficient is directly related to the renormalization 
behavior above. The basic idea behind renormalization is to use an 
effective Hamiltonian with renormalized parameters as a function of
$T$.  To estimate the reflection coefficient it is 
therefore possible to make a simplified but
intuitive analysis by using the
free fermion result in Eq.~(\ref{Rfree}), but with a renormalized
coupling  strength $\tilde J'(T)$.  
Below the cross-over temperature $T<T_K$, the effective potential 
is small and given by the renormalization behavior of the
leading irrelevant operator $\tilde J'(T) \propto J' T^{1/g-1}$. 
This results in
\begin{equation}
1-R \propto J'^2 T^{2/g-2}, \label{stable}
\end{equation}
which is the universal behavior near the stable
fixed point as first predicted in Ref.~\onlinecite{kane1}.
Above the cross-over temperature $T> T_K$ the renormalization behavior
is better described by a relevant operator on the periodic chain
giving $J - \tilde J'(T) \propto (J-J')T^{g-1}$.  
From this result it would even seem that we can recover
the periodic chain in the high temperature limit, but it is of course 
important to realize that the renormalization is no longer possible above
a cutoff of order $J$. For an initial bare coupling $J' \sim J$
very close to the unstable fixed point $T_K \ll J$
we therefore find that the effective coupling stops renormalizing 
at its bare value $\tilde J' \to J'$ for large $T$.  
In summary, the temperature dependence above $T_K$ is not as universal as
in Eq.~(\ref{stable}), but we may still write  
\begin{equation}
R \propto (J-J')^2, \label{unstable}
\end{equation}
for $J' \sim J$ and $T> T_K$.

It is now straightforward to extract the relative coefficient $c_R/c$ in 
Eq.~(\ref{linresp}) from the numerical data by simply dividing
the amplitude of the alternating part for each coupling $J'$ in
Fig.~\ref{fig:oneweakaltpart} by the reference data of $\chi^{\rm bs}$
for the open chain.
According to Eq.~(\ref{R}) the square of this relative coefficient then
gives the reflection coefficient.
Fig.~\ref{fig:Goneweak} shows the results for the temperature 
dependent reflection coefficient from our TMRG data. 
The renormalization to a perfectly
reflective barrier can clearly be seen as $T\to 0$.  
The behavior for couplings close to the periodic fixed point 
($J' \agt 0.4 J$) is consistent with 
Eq.~(\ref{unstable}). For smaller
couplings the cross-over temperature $T_K$ is larger, and we see an extended
region where the scaling of the stable fixed point with $J'^2$ and
 $T^{2/g-2}$ in Eq.~(\ref{stable}) holds (here $g=1/2$).  
We can also compare our results to the findings of
Matveev {\it et al} in Ref.~\onlinecite{yue} where an explicit formula
for the transmission coefficient was given 
$1-R \propto [(D/T)^{2 \alpha} R_0/(1 -R_0) \ + \ 1]^{-1}$
in terms of the non-interacting reflection coefficient $R_0$ in 
Eq.~(\ref{Rfree}), a cut-off $D$, and a {\it small} interaction parameter 
$\alpha = 1/g-1$.  Unfortunately, the interaction parameter  is large in
our case $\alpha = 1$ so that this formula does not quantitatively agree 
with our findings in Fig.~\ref{fig:Goneweak}.  Qualitatively,
their  result looks rather similar, but we observe a sharper renormalization
at low temperatures near the unstable fixed point ($J' \agt 0.4 J$).
Indeed we find that the region where the famous
scaling in Eq.~(\ref{stable}) is valid turns out to be extremely narrow
for $J' \agt 0.4J$.

Another aspect is the high temperature behavior where the 
non-interacting reflection coefficient in Eq.~(\ref{Rfree}) should be 
approached\cite{yue}.  This is indeed the case near the unstable fixed point
$J' \agt 0.4J$ where the non-interacting value is quickly reached
with high accuracy.  However, near the stable fixed point ($J' \alt 0.4J$)
we find that the reflection coefficient can renormalize even well below
the non-interacting value, so that the interactions actually {\it enhance}
the conductivity at higher temperatures in this case.  
The reason for this unexpected
behavior is that the cross-over temperature is larger than the cut-off
near the stable fixed point $T_K \gg J$, so that the renormalization 
may continue beyond the bare coupling constants at higher temperatures.

{\begin{figure}
\epsfxsize=7cm
\centerline{\epsfbox{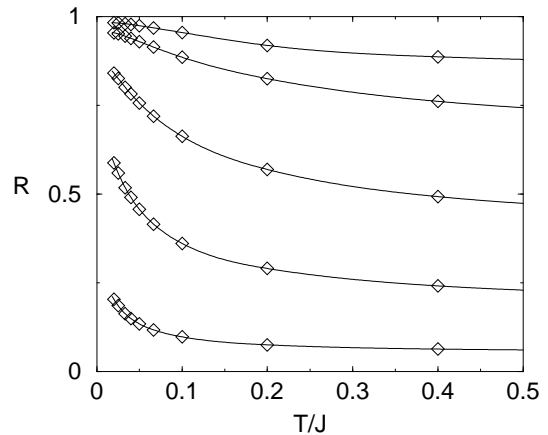}}
\caption{Reflection coefficient $R$ of one modified link for
$J'/J \ =\ 0.1, 0.2, 0.4, 0.6,
0.8$ from above. The lines are guides for the eye.}
\label{fig:Goneweak}
\end{figure}}

\subsection{Two modified links} \label{twolink}

We now consider the impurity of {\it two} neighboring 
modified links in the chain as shown in Fig.~\ref{twoweakfig}.
For the interacting case $U=2 t$ we can again write this model 
in terms of a Heisenberg spin chain model
\begin{equation}
H = J \sum_{i\neq -1,0} {\bf S}_i \cdot {\bf S}_{i+1}
 + J^\prime {\bf S}_0 \cdot \left( {\bf S}_{-1}  + {\bf
S}_1 \right). \label{twoweakham}
\end{equation}
This type of impurity may correspond to a charge island that is 
weakly coupled to a mesoscopic wire or to doping in 
a quasi-one dimensional compound where one atom in the chain
has been substituted.
We have recently considered this type of impurity in the context 
of doping in spin-1/2 compounds and as a simple experimental 
example of the two channel Kondo effect.\cite{2CK}  In this section
we analyze the induced density oscillations in more detail, especially
in connection with the reflection coefficient. 

{\begin{figure}
\epsfxsize=7cm
\centerline{\epsfbox{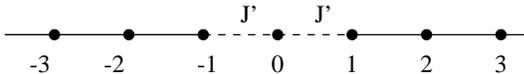}}
\caption{Two modified links.}
\label{twoweakfig}
\end{figure}}

The model in Eq.~(\ref{twoweakham}) is
equally simple as the one-link impurity, but the 
renormalization behavior is known to be quite different.\cite{eggert}
Already for the non-interacting case at half filling the system shows
a resonant behavior with perfect transmission $R=0$, so that this
corresponds to the simplest case of resonant tunneling considered
by Kane and Fisher\cite{kanereview,kane2} (at half-filling
the impurity potential is automatically tuned to the resonant condition). 
With interactions $U\neq 0$ the reflection coefficient is no longer 
exactly zero, but shows nontheless a renormalization to perfect 
transmission as $T\to 0$ in sharp contrast to the one-link impurity.
This difference in renormalization behavior is easily explained
by the different parity symmetry of the problem (namely site- instead
of link-parity).  For a small perturbation from a periodic chain $J' \sim J$
the leading operator is now {\it irrelevant} with scaling dimension of
$d=1+g$, so that a perfectly transmitting chain is the stable fixed point.
For small couplings $J' \agt 0$ on the other hand, the leading perturbing 
operator is marginally relevant, and the situation is similar 
to the two channel Kondo effect where the two ends of the chain play the
role of two independent channels.\cite{eggert,2CK}  

Apart from the renormalization behavior there is another key difference
between the one- and two-link impurities:  In the two-link impurity model
there is an ``active'' impurity site that carries a spin or charge degree
of freedom, which in turn must be {\it screened} by the surrounding system.
Therefore, the density oscillations are no longer simply determined by 
the backscattering in Eq.~(\ref{chi}), but there is also a so-called
screening cloud induced in the system.  From perturbation theory in the 
leading irrelevant operator the functional dependence of this screening
cloud can be calculated\cite{2CK} and the total alternating density
$\chi^{\rm alt}$ is a sum of two contributions
\begin{equation}
\chi^{\rm alt} (x) = c_I (-1)^{x} \ln[\coth(x T)] - 
c_R (-1)^{x} \chi^{\rm bs}(x),
\label{twolinkalt} 
\end{equation}
where the first term is the induced screening cloud while the second
term is the familiar contribution due to backscattering in Eq.~(\ref{chi}).
Interestingly, the two contributions have opposite sign, so that 
the density oscillations vanish at a special distance from the
impurity, but then increase again due to the backscattering contribution.  
This behavior is shown in 
Fig.~\ref{fig:altchitwoweak} together with a fit to the two contributions
in Eq.~(\ref{twolinkalt}).
The special distance at which the density oscillations vanish grows
as we approach the stable fixed point ($J' \to J$ or $T \to 0$).
As already with the one-link problem, we use again the numerical 
open chain data as a reference for $\chi^{\rm bs}$
instead of the more simplified analytical form of the
backscattering contribution in Eq.~(\ref{chi}) since this minimizes 
the corrections due to irrelevant operators. However, even the 
analytical form in Eq.~(\ref{chi}) gives very good fits, so that none of our 
our findings are affected by this choice.

{\begin{figure}
\epsfxsize=7cm
\centerline{\epsfbox{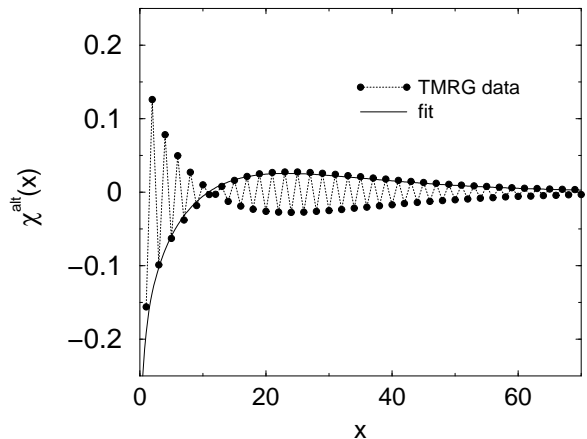}}
\caption{Alternating part of the local susceptibility for the two-link
impurity for $T/J=0.04$ and $J^\prime/J=0.6$. Fit to
Eq.~(\ref{twolinkalt}).}
\label{fig:altchitwoweak}
\end{figure}}

It is now straightforward to extract the reflection coefficient
from the numerical data 
with the help of Eq.~(\ref{R}) and Eq.~(\ref{twolinkalt}) as shown in 
Fig.~\ref{fig:Gtwoweak}.  Below a cross-over temperature $T_K$ 
depending on $J'$
the reflection coefficient clearly decreases and eventually 
approaches perfect transmission as $T\to 0$. 
Above $T_K$ the renormalization of the reflection coefficient is
rather weak and converges to a finite constant (never approaching 
complete reflection as the temperature increases).
 
{\begin{figure}
\epsfxsize=7cm
\centerline{\epsfbox{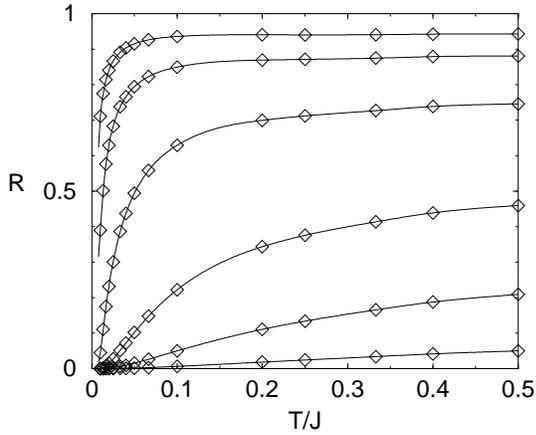}}
\caption{Reflection coefficient $R$ of the two-link impurity for
 $J'/J = 0.05, 0.1, 0.2, 0.4, 0.6,
0.8$ from above. The lines are guides for the eye.}
\label{fig:Gtwoweak}
\end{figure}}

Equally interesting is the induced screening cloud.  In this case,
the coefficient $c_I$ approaches a constant as $T < T_K$ as it should,
since this contribution was determined from perturbation theory around
the stable fixed point.  Above the cross-over temperature, however, 
this contribution vanishes quickly. This behavior is shown in
Fig.~\ref{fig:c1twoweak}:  In general the behavior of the 
coefficient $c_I$ vs. $J'$ is temperature dependent and $c_I$ increases as
the temperature is lowered.  However, as $T \ll T_K$
all curves approach a limiting value, which gives a universal behavior 
as a function of $J'$ (thick line).

{\begin{figure}
\epsfxsize=7cm
\centerline{\epsfbox{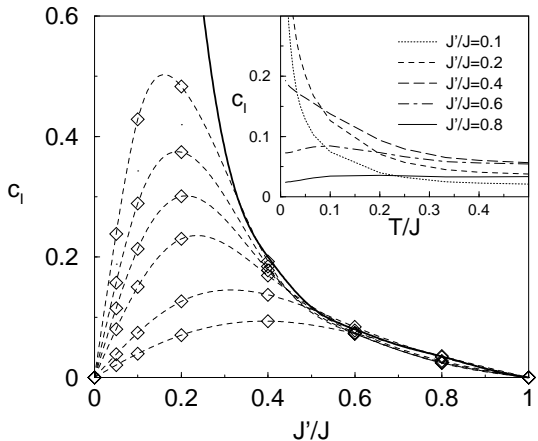}}
\caption{Coefficient $c_I$ vs $J'$ of the two link impurity
for different temperatures $T/J =$
$0.2,0.1,0.04,0.025,0.0167,0.01$
from below. For $J' \approx J$ and/or low temperatures
$c_I$ approaches a universal $T$-independent curve (thick line).
Inset: $c_I$ vs $T$.  The lines are guides for the eye.}
\label{fig:c1twoweak}
\end{figure}}
 
The competing contributions in Eq.~(\ref{twolinkalt})
have the opposite renormalization behavior:
Above $T_K$ backscattering is constant, while the screening cloud is
reduced which is the open chain behavior. Below $T_K$ on the other hand
backscattering is reduced, while the coefficient for the 
induced screening cloud is constant, which is the behavior of the two channel 
Kondo fixed point.  Note, that although the coefficient $c_I$ is finite as
$T\to 0$, the screening cloud itself diverges logarithmically with
$-\ln(x T)$, which is a clear indication of the famous over-screening
in the two channel Kondo effect.  As we approach the unstable fixed point 
the order of limits becomes crucial:  
For zero coupling there is no screening cloud at all
$\lim_{T\to 0} \lim_{J' \to 0} c_I = 0$,  while
for zero temperature the
coefficient becomes infinite $\lim_{J'\to 0} \lim_{T \to 0} c_I = \infty$. 
Remarkably, exactly at zero 
temperature a minute perturbation therefore induces an infinite screening
cloud, although this behavior occurs in an unphysical limit.


\subsection{Impurity at the edge}  \label{edgeimp}

Another category of impurities we can consider are imperfections
near the end of a chain.  In this case the boundary always gives
complete backscattering, but as we will see the impurity 
can still give interesting effects on the density oscillations.
The simplest case to consider is a modified link at the edge of
a chain as depicted in Fig.~\ref{fig:halfedge}.  For the interacting
case $U=2 t$ it is again useful to write the Hamiltonian in terms
of the Heisenberg spin-chain model
\begin{equation}
H = J \sum_{i=1}^\infty {\bf S}_i \cdot {\bf S}_{i+1}
 + J^\prime {\bf S}_0 \cdot {\bf S}_1 .
\label{edgeham}
\end{equation}

{\begin{figure}
\epsfxsize=7cm
\centerline{\epsfbox{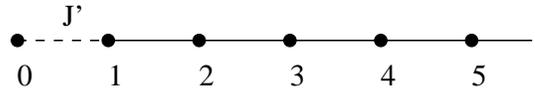}}
\caption{Edge impurity.}
\label{fig:halfedge}
\end{figure}}

Just like the two-link impurity was related to the two-channel
Kondo problem, we can identify the field theory description of the
edge impurity model with the regular
one-channel Kondo problem.  There are two possible fixed points:
The case $J' = 0$ corresponds to the unstable fixed point of a decoupled
spin at the end of a chain with a marginally relevant perturbation
for $J' \agt 0$.  The case $J' = J$ corresponds to  the completely 
screened spin, which is a stable fixed point with a 
leading irrelevant operator of scaling dimension $d=2$. 
Just like in the ordinary Kondo effect both fixed points 
are represented by the same boundary condition and differ only
by a simple $\pi/2$ phase shift on the fermions.  (The infinite
coupling fixed point $J' \to \infty$ is also stable, but is actually 
absolutely
equivalent to the $J'=J$ fixed point since both cases represent a $\pi/2$
phase shift on the fermions by removing or adding a site, respectively).
For intermediate
couplings the phase shift $\Phi$ takes on values between 0 and $\pi/2$ which 
will be reflected in the backscattering contribution of the density
oscillations as we will see below.

A screening cloud for the impurity spin at the end should
also be present in this model, but with a different behavior than
for the overscreened case in Eq.~(\ref{twolinkalt}).  Instead we find that
the leading operator that causes the screening cloud is the same 
as that for an edge magnetic field in the xxz-chain which has been analyzed in 
Ref.~\onlinecite{affleckedgefield}, so we can use the corresponding
result for the shape of the induced screening cloud.  
Taking into account finite temperatures and the phase shift 
on the fermions we can write for the density oscillations
\begin{equation}
\chi^{\rm alt}(x) = c_I \frac{(-1)^x \sqrt{T}}{\sqrt{\sinh(4xT)}}  -
\cos(\pi x +2\Phi) \ c \ \chi^{\rm bs}(x), \label{edgealt}
\end{equation}
where the first term is the induced screening cloud, while the second
term is the backscattering contribution in Eq.~(\ref{chi}) but
with a phase shift $\Phi$.  
However, the coefficient $c$ always takes the value corresponding to complete 
backscattering in Eq.~(\ref{susc}).  There is also an implied shift of 
$2 \Phi/\pi$ in the argument of $\chi^{\rm bs}$, which we used
for a self-consistent fitting.  The effective boundary condition in the
continuum limit is therefore technically between two lattice sites (although
it is not really that meaningful to define locations on the scale of less than
a lattice spacing in the continuum limit theory anyway).

Figure~\ref{fig:Szhalfedge} shows the envelope of the alternating part of
the susceptibility for temperature $T=0.04J$ and different couplings $J'$,
which always fits well to the superposition in Eq.~(\ref{edgealt}). At the
fixed points $J'=0$ and $J'=J$ there is no screening, but the backscattering
contribution has opposite signs due to the $\pi/2$ phase shift.

{\begin{figure}
\epsfxsize=7cm
\centerline{\epsfbox{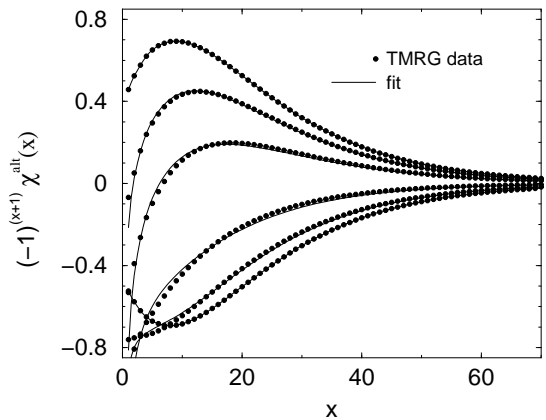}}
\caption{Alternating susceptibility for the edge impurity at $T/J=0.04$  for
$J'/J = 0, 0.1, 0.2, 0.4, 0.6, 1.0$ from above. Fits to Eq.~(\ref{edgealt}).}
\label{fig:Szhalfedge}
\end{figure}}

It is now straightforward to extract the screening cloud amplitude $c_I$
and the phase shift $\Phi$ from our numerical data for all temperatures
and couplings $J'$. As expected we find that the phase shift increases with
$J'$ and renormalizes to larger 
values as the temperature is lowered as shown in Fig.~\ref{fig:phihalfedge}.
In the limit of very low temperatures the jump to the stable fixed point value 
$\Phi =\pi/2$ becomes more abrupt as a function of $J'$.  

{\begin{figure}
\epsfxsize=7cm
\centerline{\epsfbox{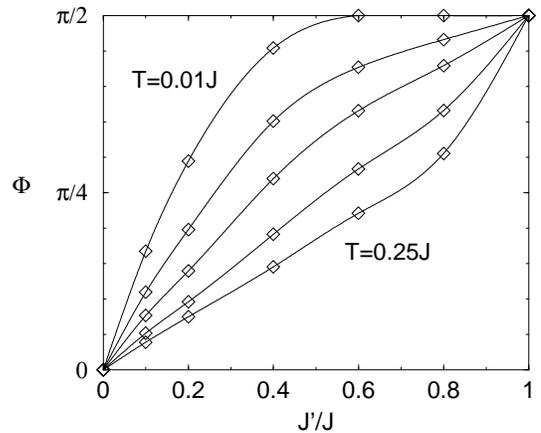}}
\caption{Phase shift of alternating part of the edge impurity for
$T/J = 0.25, 0.1, 0.04, 0.02, 0.01$ from below. The lines are guides
for the eye.}
\label{fig:phihalfedge}
\end{figure}}

The screening cloud coefficient $c_I$ again approaches a constant as we 
lower the temperature below $T_K$ as shown in Fig.~\ref{fig:c1halfedge}.
Although formally the behavior looks similar to the 
over-screened case of the two link problem in Fig.~\ref{fig:c1twoweak}
it is important to realize that now the screening cloud in 
Eq.~(\ref{edgealt}) is finite as $T\to 0$ and drops off with $1/x$ (while
in the two link case the screening cloud was divergent with $\ln xT $).

{\begin{figure}
\epsfxsize=7cm
\centerline{\epsfbox{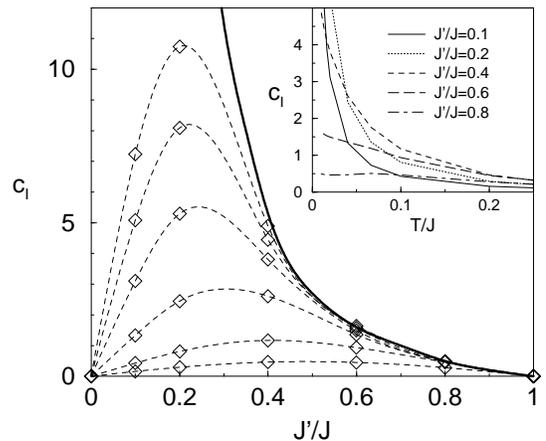}}
\caption{Coefficient $c_I$ vs. $J'$ of the edge impurity for
$T/J = 0.2,0.1,0.04,0.02,0.133,0.01$ from below.
The lines are a guide for the eye.}
\label{fig:c1halfedge}
\end{figure}}

\subsection{Generalized two link impurity} \label{genimp}

It is now instructive to summarize the findings of the three
impurity models in the previous subsections by considering one
generalized two link impurity model that is {\it not symmetric}
as shown in Fig.~\ref{genfig}
\begin{equation}
H = J \sum_{i\neq -1,0} {\bf S}_i \cdot {\bf S}_{i+1}
 + J_1 {\bf S}_{-1} \cdot  {\bf S}_{0} + J_2 {\bf S}_0 \cdot   {\bf
S}_1 . \label{genham}
\end{equation}
The three impurity cases above can be identified easily:
\begin{itemize}
\item $J_2 \neq J_1 = J$  one modified link in Eq.~(\ref{oneweakham})
\item $J_1 = J_2 \neq J$  two modified links in Eq.~(\ref{twoweakham})
\item $J_1 = 0, \  J_2 \neq J$ edge impurity in Eq.~(\ref{edgeham})
\end{itemize}
The density oscillations for the more general model in Eq.~(\ref{genham})
are much more complex than in the special cases, so that a detailed
analysis of this effect is not always useful.  The renormalization behavior
on the other hand is straightforward and can be read off from what we 
already know about the special cases.

{\begin{figure}
\epsfxsize=7cm
\centerline{\epsfbox{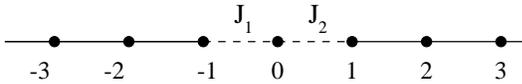}}
\caption{Generalized two-link impurity.}
\label{genfig}
\end{figure}}

A weak coupling $J_1 \agt 0$ and $J_2 \agt 0$ to an additional site
is always marginally relevant, so that the open chain with a 
decoupled impurity site is unstable for any antiferromagnetic coupling
(i.e.~negative hopping probability). 
The periodic chain on the other hand is only stable for the special
site-parity symmetric case $J_1 = J_2$, where the 
renormalization behavior is analogous to the two channel Kondo 
effect.  In general, however, one of the two couplings is larger
and renormalizes to unity, absorbing the spin.  The smaller coupling
is then irrelevant as in the one-weak problem, so that the stable
fixed point is an open chain with an absorbed impurity 
site $J_1 = J, \ J_2 =0$ (or $J_2 = J, \ J_1 =0$) in most cases, 
except for a site-parity symmetric impurity  
or two ferromagnetic coupling constants.  The complete renormalization 
flow is summarized in Fig.~\ref{renflow} where the possible 
fixed points are indicated by the black dots.  In cases where the
coupling diverges to infinity a singlet forms, and we can therefore again
describe the system by one of the four finite fixed points in
the figure.  Interestingly, the more stable fixed points
always have a lower ground state degeneracy, in accordance with the 
g-theorem.\cite{g-theorem}  The phase diagram in Fig.~\ref{renflow}
is valid for all interaction strengths $0 < U \leq 2 t$ as long as
the system is half-filled.
 
{\begin{figure}
\epsfxsize=7cm
\centerline{\epsfbox{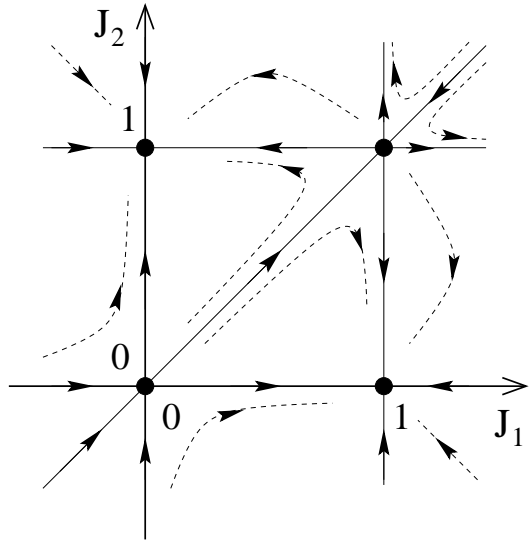}}
\caption{Renormalization flow diagram.}
\label{renflow}
\end{figure}}
 
\subsection{Spin-1 impurity} \label{spin1}

We now turn to a magnetic impurity in the chain with spin $S_{\rm imp}=1$ 
given by the Heisenberg Hamiltonian
\begin{equation}
H =  J \sum_{i\neq 0} {\bf S}_i \cdot {\bf S}_{i+1}
 + J^\prime {\bf S}_{\rm imp} \cdot \left( {\bf S}_{0}  + {\bf
S}_1 \right). \label{spin1ham}
\end{equation}
as shown in Fig.~\ref{S1_twoweakfig}.
In the previous impurity models in 
Sections \ref{onelink}-\ref{genimp}
it was always possible to interpret 
the Heisenberg Hamiltonians equally well in terms of mesoscopic 
systems and electrons 
hopping on the lattice by identifying the spin-1/2 impurity
in terms of an extra site or charge island.  However, for the 
spin-1 impurity in Eq.~(\ref{spin1ham}) no meaningful
interpretation in terms of spinless fermions is possible. On the
other hand this impurity model has important implications for doping in 
quasi one-dimensional spin-1/2 compounds, so that we find it useful to 
discuss it here.

{\begin{figure}
\epsfxsize=7cm
\centerline{\epsfbox{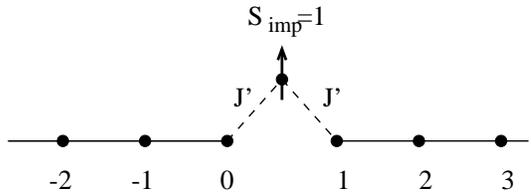}}
\caption{The spin-1 impurity.}
\label{S1_twoweakfig}
\end{figure}}

Similar to the impurity models in Sections \ref{twolink} and \ref{edgeimp}
we find again that the field theory language is analogous to a
Kondo impurity model.  The two ends of the spin-chain play the role
of the two channels coupled to a spin-1 impurity.  A small antiferromagnetic
coupling is therefore marginally relevant and the renormalization
flow goes to the strong coupling limit. The stable fixed point is 
given by an open spin chain with two sites removed and a decoupled singlet 
containing the spin-1 and the two end spins ($J'\to \infty$).

Just like the edge impurity in Sec.~\ref{edgeimp} this Kondo-type model 
is an exactly screened impurity. The shape of the screening cloud
is again given by that of an edge magnetic field\cite{affleckedgefield}
just like in Eq.~(\ref{edgealt})
\begin{equation}
\chi^{\rm alt}(x) = c_I \frac{(-1)^x \sqrt{T}}{\sqrt{\sinh(4xT)}} - 
 c_R (-1)^{x} \chi^{\rm bs}(x)
, \label{spin1alt}
\end{equation}
where the first term is again the induced screening cloud, while the second
term is the backscattering contribution in Eq.~(\ref{chi}). As shown in
Fig.~\ref{fig:S1_twoweakc1} the fits to this expression are 
excellent (again using the open chain data as a reference for $\chi^{\rm bs}$).
The coefficient $c_I$ for the induced screening cloud again approaches a 
constant for temperatures below $T_K$ which results in a universal
curve as $T\to 0$ as shown in Fig.~\ref{fig:S1_c1twoweak}.
The backscattering coefficient is an indication of the 
effective phase shift and changes sign depending 
on the temperature and coupling strength.  
From Fig.~\ref{fig:S1_twoweakc1}
it is clear that the backscattering coefficient $c_R$ is positive
for small coupling strengths $J'$ (or equivalently high temperatures) 
and negative for
larger coupling strengths $J'$ (or equivalently lower temperatures).
The renormalization of $c_R$ is explicitly shown in the inset of 
Fig.~\ref{fig:S1_c1twoweak}.  As $T\to 0$ the jump of $c_R$ to negative
values happens at smaller $J'$ and becomes very sharp.

{\begin{figure}
\epsfxsize=7cm
\centerline{\epsfbox{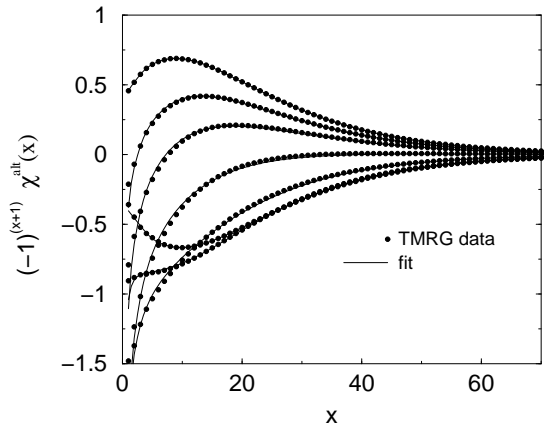}}
\caption{Envelope of alternating part at $T/J=0.04$ for the
spin-1 impurity. From above at $x \protect\agt 20$:
$J^\prime/J = 0, 0.05, 0.1, 0.2, 0.4, 0.8, 4.0$.
Fits to Eq.~(\ref{spin1alt}).}
\label{fig:S1_twoweakc1}
\end{figure}}

{\begin{figure}
\epsfxsize=7cm
\centerline{\epsfbox{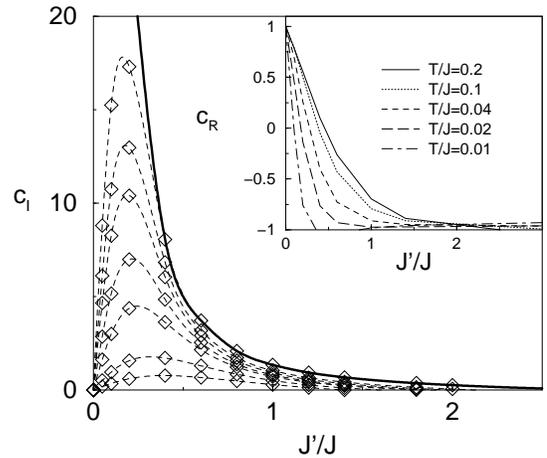}}
\caption{Coefficient $c_I$ of the spin-1 impurity for
$T/J =$
$0.2,0.1,0.04,0.025,0.0167,0.0133,0.01$ from below. Inset: Backscattering
coefficient $c_R$.  The dashed lines are a guide for the eye.}
\label{fig:S1_c1twoweak}
\end{figure}}

More interesting are the experimental consequences for Knight shift 
experiments in doped spin-1/2 chain compounds (as for example
Ni doping in CuO chains).  For that case we can predict an interesting NMR
spectrum with a characteristic feature (sharp edge) corresponding to the 
maximum in the alternating susceptibility.  Such a sharp edge has been 
observed before in NMR experiments on spin-1/2 chain compounds with
non-magnetic defects.\cite{NMR}  In that case the sharp edge broadens with
a $1/\sqrt{T}$ behavior as discussed in Sec.~\ref{model}.
For the magnetic spin-1 impurities a sharp edge from the maximum in the 
backscattering part may also be present, but it depends on if the temperature 
is above or below $T_K$ how this feature changes.
Above $T_K$ the backscattering part becomes weaker as the temperature 
is lowered,
but the induced screening cloud increases, so that the sharp kink 
may vanish in a quickly broadening line-shape from the screening cloud
as shown in the left part of Fig.~\ref{fig:S1_twoweakNMR2}.
Below $T_K$ on the other hand, the screening has saturated and
the backscattering contribution dominates again (albeit with
a phase shift).  Therefore, the kink feature in the NMR spectrum will
sharpen further as the temperature is lowered and widen with the usual 
$1/\sqrt{T}$ behavior as shown in the right part 
of Fig.~\ref{fig:S1_twoweakNMR2}.  The detailed T-dependence can be
predicted for any particular value of $J'$ of an actual experimental
compound.

{\begin{figure}
\epsfxsize=7cm
\centerline{\epsfbox{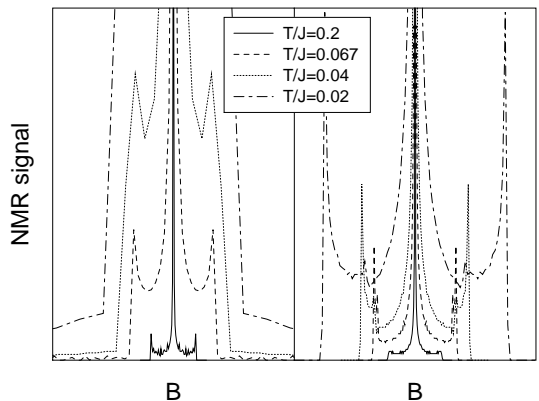}}
\caption{NMR signal of the spin-1 impurity for $J^\prime/J=0.1$
(left) and $J^\prime/J=1.4$ (right). }
\label{fig:S1_twoweakNMR2}
\end{figure}}

\subsection{Integrable impurity model}

Finally, we would like to consider a more exotic impurity model 
which has been especially constructed to preserve the integrability 
of the entire system.\cite{henrik}  We consider here the simplest
non-trivial example of such an impurity model which corresponds
to an impurity spin with $S_{\rm imp} = 1$
that is coupled in a special way to two
sites in the chain.  The corresponding Hamiltonian has been set up 
in Ref.~\onlinecite{henrik}
\begin{eqnarray}
H & = &
 J \sum_{i\neq 0} {\bf S}_i \cdot {\bf S}_{i+1} 
- \case{7 J}{9} {\bf S}_{0} \cdot {\bf S}_1 
\label{intham} \\
& & + \case{4 J}{9} 
\left[ ({\bf S}_{0}  +
 {\bf S}_1) \cdot {\bf S}_{\rm imp} 
+ \left\{ {\bf S}_{0} \cdot {\bf S}_{\rm imp}, {\bf S}_1 \cdot {\bf
S}_{\rm imp} \right\} \right] , 
\nonumber 
\end{eqnarray}
where ${\bf S}_{\rm imp}$ is the external spin-1 impurity
and $\{,\}$ denotes the anticommutator. 

{\begin{figure}
\epsfxsize=7cm
\centerline{\epsfbox{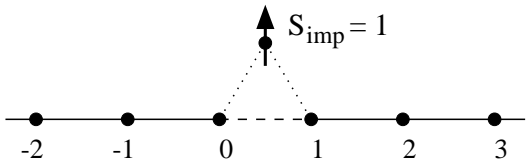}}
\caption{The integrable impurity.}
\label{fig:integrable}
\end{figure}}

A closer analysis of this model\cite{schlottmann} 
showed that the thermodynamics
at low temperatures were in fact described by a {\it periodic}
spin chain with one additional site and an asymptotically free impurity
spin with $S=1/2$, so that it appears that the original spin-1 has
somehow been partially absorbed by the chain.  From a field theory
point of view it was later shown that this type of impurity corresponds 
in fact to an unstable fixed point which can only be reached by an artificial
tuning of the coupling parameters.\cite{erik}

We are now interested in what kind of density oscillations might be
observable from such an impurity.  Interestingly, we found that the
density oscillations were {\it identically zero at all temperatures} 
as if the system was translationally invariant.
The impurity Hamiltonian in Eq.~(\ref{intham}) was of course constructed
in a way to avoid all backscattering, but it is remarkable that even 
the induced alternating part from the magnetic impurity vanishes exactly, 
i.e.~no conventional screening takes place.

Nonetheless, the impurity spin is somehow reduced from a spin-1 to an
effective spin-1/2 as the temperature is lowered.  This can be explicitly
seen from the impurity susceptibility
in small magnetic fields  
\begin{equation}
\langle S^z_{\rm imp}\rangle = B\frac{C_{\rm Curie}}{T}
\end{equation}
where we have assumed some type of Curie-law.
At high temperatures
the impurity susceptibility must follow the Curie-law for a spin-1
$C_{\rm Curie} = 2/3$, 
while at low temperatures a Curie-law for a spin-1/2 $C_{\rm Curie} = 1/4$
has been predicted up to logarithmic corrections.\cite{schlottmann}  
In Fig.~\ref{fig:chiintegrable}
we plot the temperature dependent Curie constant (i.e. the
impurity susceptibility times temperature).  It appears that the
asymptotic value $C_{\rm Curie} = 1/4$ is indeed approached with 
logarithmic corrections as $T\to 0$. The fit in the figure is 
\begin{equation}
C_{\rm Curie} = 
\frac{1}{4} + \frac{1}{8 \ln(2 \pi/T)} + a \frac{\ln \left( \ln(2 \pi/T)/b
\right) }{\ln(2 \pi/T)^2} \label{integrfit}
\end{equation}
with $a=1.62$ and $b=1.32$.

{\begin{figure}
\epsfxsize=7cm
\centerline{\epsfbox{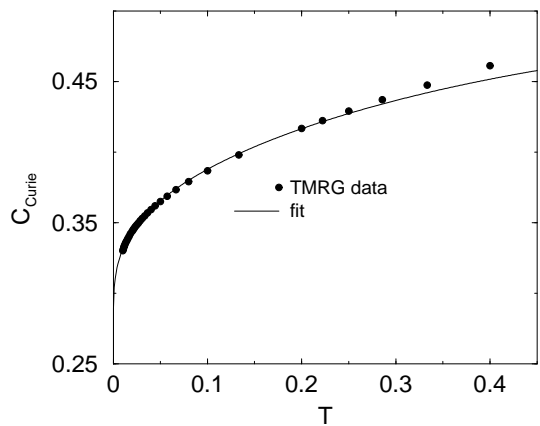}}
\caption{Susceptibility of the spin-1
in the integrable model multiplied by temperature. Fit to
Eq.~(\ref{integrfit}).}
\label{fig:chiintegrable}
\end{figure}}


\section{Numerical Method} \label{numerics}
The numerical method we have 
used here is based on the Density Matrix Renormalization
Group (DMRG)\cite{DMRG} applied to transfer matrices.  
While the ordinary DMRG considers the properties of individual 
eigenstates in a finite system, we are interested in the thermodynamic
limit, namely properties of an infinite system at finite temperatures.
This can be achieved by the Transfer Matrix Renormalization Group 
(TMRG),\cite{TMDMRG} which we adapted especially for 
impurities\cite{2CK,rommer} as we will review briefly.
We consider the partition function $Z$ of the
models in Eqs.~(\ref{fermions}) and (\ref{heisenberg}). After the 
standard Trotter decomposition, we obtain for an infinite 
system $(L\to \infty)$
\begin{equation}
Z = \lim_{M\to \infty} {\rm tr} T_M^{L/2} \to 
\lim_{M\to \infty} \lambda_M^{L/2}, \label{Z}
\end{equation}
where $T_M$ is the transfer matrix with $M$ time-slices. In the limit of 
infinite system size only the largest eigenvalue $\lambda_M$ determines
the thermodynamics of the system, which we find numerically.
We start with small time-steps so that the Trotter-error is negligible, and
successively increase the number of time-slices $M$ to reach lower 
temperatures.  At each step the dimension of $T_M$ increases so we keep only
the most important states to describe the state with the highest 
eigenvalue $\lambda_M$ by using the DMRG algorithm with some modifications
for asymmetric matrices.\cite{rommer}
A measurement of the local spin-density at site $j$
for example is straightforward,
since we can just absorb the measuring operator 
$S^z_j$ into one of the transfer matrices $T_M \to T^{sz}_M$
\begin{equation}
\langle S_j^z \rangle = \frac{1}{Z} \; \mbox{tr} \; S_j^z e^{-\beta H}
\to \frac{\langle \psi_M | {T}_M^{sz}(j) | \psi_M \rangle}{\lambda_M},
\label{sz}
\end{equation}
where $\langle \psi_M |$ and $| \psi_M \rangle$ are the left and right 
target states for the eigenvalue $\lambda_M$.  So far we have considered
a translational invariant system.

We now introduce a generic impurity which modifies one of the
transfer matrices $T_M \to T_{\rm imp}$.  
Even in the presence of impurities the thermodynamics of the system
is entirely determined by the highest eigenvalue $\lambda_M$ and 
corresponding eigenstate of the pure transfer matrix $T_M$ which always 
appears with an infinite power in the partition function in Eq.~(\ref{Z}).
The measurement of
the spin (or charge) density near the impurity is again straightforward.
For the spin density at a distance of $j$ sites from the 
impurity we write
\begin{equation}
\langle S^z_j \rangle = \frac{ \langle \psi_M | {T}_{M}^{sz} 
\ ({T}_{M})^{j/2} \ {T}_{\rm imp} | \psi_M \rangle }{ \lambda_M^{j/2+1} \;
\langle \psi_M | {T}_{\rm imp} | \psi_M \rangle } . \label{szimp}
\end{equation}

Since we step-wise approximate the transfer matrix, it is
important to make a careful error-analysis.  The error due to the
Trotter approximation is the simplest to estimate since it
is just proportional to the square of the time-step $\tau = 1/T M$.
We found that a value of $\tau = 0.05/J$ makes this error negligible
compared to the DMRG truncation errors.  To estimate the truncation
errors we can compare our results to the exact solution of the
free fermion Hamiltonian in Eq.~(\ref{fermions}) with $U=0$.
The structure of the transfer matrix is not fundamentally changed
by taking $U=0$ so that the truncation error will be of the same 
order as for $U\neq 0$. Keeping 64 states we find
for the local response of the spins closest to
typical impurities a relative error of less than
$10^{-4}$ for $T>0.04$, less than $10^{-3}$ for $0.02 < T < 0.04$ and a relative
error of less than $10^{-2}$ for temperatures $0.01 < T < 0.02$. However, already
from Eq.~(\ref{szimp}) it is clear 
that the spin and charge densities far away from the impurity will
contain a larger error.  Each transfer matrix contains a small
error $\epsilon$ which then gets exponentiated in Eq.~(\ref{szimp})
and hence the oscillating part
of the density $\langle S_j^z \rangle$ is suppressed exponentially with
distance $j$ 
\begin{equation}
\langle S_j^z  \rangle_{\rm osc} \propto (1-\epsilon)^{j} = \exp(-j \epsilon)
\label{error}
\end{equation}
where $\epsilon$ depends only on temperature.  This exponential 
suppression with the distance from the boundary
is again a consequence of the fact that the incoming and
outgoing waves lose coherence but this time 
due to error fluctuations.  However, the corresponding energy 
scale from the truncation error is always smaller than the 
temperature in our case.  We observe that the suppression error 
in Eq.~(\ref{error}) is actually very systematic, so that we 
can even correct our data very well using Eq.~(\ref{error}).
For free fermions we find to high accuracy the following dependence
of the error on temperature
\begin{equation}
\epsilon = 0.06 \exp(- 58 T),
\label{epsilon}
\end{equation}
where we have kept 64 states in the TMRG simulations.
For interacting fermions the suppression also has the exponential dependence
in Eq.~(\ref{error}), but the energy scale $\epsilon$ is in general
dependent on the interaction $U$.  For the Heisenberg model an 
independent analysis of the free energy hinted at a value of 
approximately $\epsilon = 0.02 \exp(-34 T)$, but the value in 
Eq.~(\ref{epsilon}) is more reliable and gives a relatively 
good estimate of the error for all interaction strengths.
We chose to correct our data for the alternating 
fermion densities by dividing out the factor in Eq.~(\ref{error})
together with the estimate in Eq.~(\ref{epsilon}) in all cases presented
above.  However, the use of this correction or the 
particular choice of the error $\epsilon$ makes no qualitative difference
in any of our findings, since the temperature suppression always 
dominates (i.e.~the energy scale in Eq.~(\ref{epsilon}) is always
smaller than the temperature).  
Another important energy scale is the finite magnetic field $B$
that is used in the simulations (i.e.~how close the system is to 
half filling).  We typically used a value of $B=0.003$ which makes the 
magnetic length scale in Eq.~(\ref{n_osc2})
always negligible compared to the finite temperature correlation length.

\section{Conclusion} \label{summary}
We have considered a number of impurity models and were able to 
extract detailed information about the backscattering amplitude,
the backscattering phase-shift, and the impurity screening effects
by examining the Friedel oscillations.  The results for the
various impurities have direct and indirect implications for
a large number of theoretical models and experimental systems
as we will summarize below.

\subsection{Kondo-type impurities}
Kondo impurity problems are maybe the most famous examples of
impurity renormalization effects ever since the classic work by 
Wilson.\cite{wilson}  
Many of the impurity models we have considered here are analogous to 
Kondo impurity problems in terms of the field theory language. 
In particular, the field theory description of a Heisenberg chain 
is the same as that of the spin-channel for a spin-full electron
field (while the charge excitations are neglected).
Moreover, it is known that coupling the open end of a Heisenberg chain
to an impurity spin produces the same impurity operators as in
the real Kondo problem.\cite{eggert,affleck} 
The number of channels in the equivalent Kondo problems 
is given by the open ends that the 
impurity spin is connected to (e.g.~the two link impurity in 
Sec.~\ref{twolink} is analogous to the two channel S=1/2 Kondo problem).
It is important to realize that the Heisenberg spins 
in the chains that we consider here have
different expressions in terms of the boson fields than the real 
electron spins in the full three dimensional Kondo problems.
Nonetheless, we can still use our models to gain some insight into the 
central aspects of renormalization, scaling, cross-over temperature, and
screening clouds.  

We have shown that the Kondo-type impurities indeed show the 
expected renormalization to a screened impurity spin. In particular, 
we have found a diverging screening cloud (and vanishing backscattering)
for the overscreened case in Sec.~\ref{twolink}, while the 
exactly screened cases in Sec.~\ref{edgeimp} and \ref{spin1}
are characterized by a finite screening cloud and a phase shift in
the backscattering as $T\to 0$.  

To analyze the renormalization process more quantitatively it is important
to introduce the concept of scaling.  It can be expected that 
the impurity introduces a new energy scale that depends on the initial bare 
coupling constants.  Commonly this energy scale is 
referred to as the cross-over temperature $T_K$.  By making use of 
scale invariance it is then possible to describe the renormalization
process universally in terms of the single parameter $T/T_K$.
In particular, impurity properties like the impurity susceptibility 
are described by a universal scaling function
$\chi_{\rm imp} = f(T/T_K)/T$,  which is valid for all $T$ and $T_K$ below
the cut-off.  This behavior was demonstrated explicitly before 
for the two weak link problem\cite{2CK} and works for all
Kondo-type impurities in this paper (not shown).  In fact it is 
possible to extract the cross-over 
temperature $T_K$ up to an arbitrary overall scale explicitly
by collapsing the data according to the scaling analysis.\cite{2CK,rommer}
We have determined $T_K$ this way as a 
function of coupling $J'$ in each case as
shown in Fig.~\ref{fig:tkcollapse} (up to an arbitrary overall scale).
The Kondo temperature shows the same exponential dependence for small $J'$ 
\begin{equation}
T_K \propto  \exp(-0.85J/J') \label{TK}
\end{equation}
as shown in Fig.~\ref{fig:tkcollapse} (coming from the same
marginally relevant operator at the unstable fixed point in all cases).
The underscreened case of a spin-1 coupled to the end of one chain has also 
been included in Fig.~\ref{fig:tkcollapse} for completeness.

{\begin{figure}
\epsfxsize=7cm
\centerline{\epsfbox{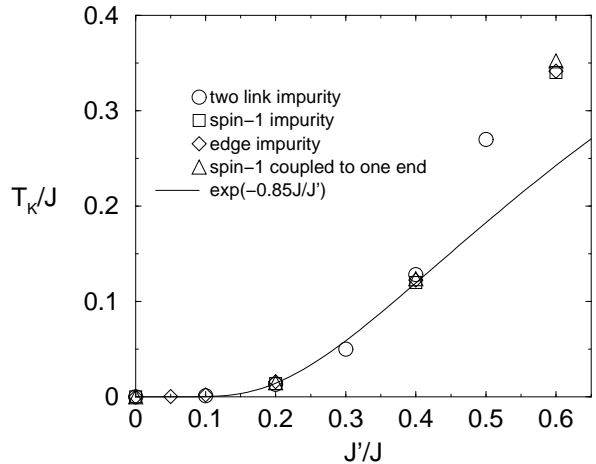}}
\caption{Crossover temperature $T_K$ of four different Kondo-type impurities.
$T_K$ has been multiplied by arbitrary constants in order to compare
the four cases.}
\label{fig:tkcollapse}
\end{figure}}

More interesting in the context of the density oscillations
is maybe the scaling of the screening cloud.  As the screening cloud
we define that part of the alternating density that is induced by
the magnetic impurity, labeled by $c_I$ in Eqs.~(\ref{twolinkalt}), 
(\ref{edgealt}), and (\ref{spin1alt}).  In Ref.~\onlinecite{sorensen}
it was postulated that the screening cloud in the real
Kondo effect should be a function of the 
scaling variables $xT$ and $T/T_K$.
In our cases we can make a similar argument except that we need
to include an overall factor $T^{g-1}$ to account for the dimensionality 
of the correlation functions.  We therefore obtain the 
following scaling law
\begin{equation}
\chi^{\rm screening} = T^{g-1} f(x T, T/T_K). \label{scaling}
\end{equation}
Indeed we find that 
the shape of the screening cloud is not affected by $T_K$
and can always be expressed as a function of the scaling variable $xT$.  
The coefficient $c_I$
must therefore be a function of $T/T_K$ multiplied by 
appropriate powers of $T$. As an example we can take the
two link problem at $g=1/2$ with the screening cloud given in 
Eq.~(\ref{twolinkalt}), where the coefficient can be written as
$c_I = f(T_K/T)/\sqrt{T}$ with some function $f$.  In Fig.~\ref{cIscale}
we replot the coefficient $c_I$ analogous 
to Fig.~\ref{fig:c1twoweak} but with the argument replaced by $T_K/T$
instead of $J'$.  The inset shows that  
the data indeed collapses if multiplied by $\sqrt{T}$ as implied by 
Eq.~(\ref{scaling}).  The solid line in Fig.~\ref{fig:c1twoweak}
therefore is proportional to $1/\sqrt{T_K}$ and diverges exponentially
with  $J'$ according to Eq.~(\ref{TK}).
Similar arguments can be made for the coefficients $c_I$  in the screening 
clouds of the exactly screened cases in Eqs.~(\ref{edgealt}) 
and (\ref{spin1alt}), except that $c_I = f(T_K/T)/T$ 
and the solid line is proportional to $1/T_K$ in that case.

{\begin{figure}
\epsfxsize=7cm
\centerline{\epsfbox{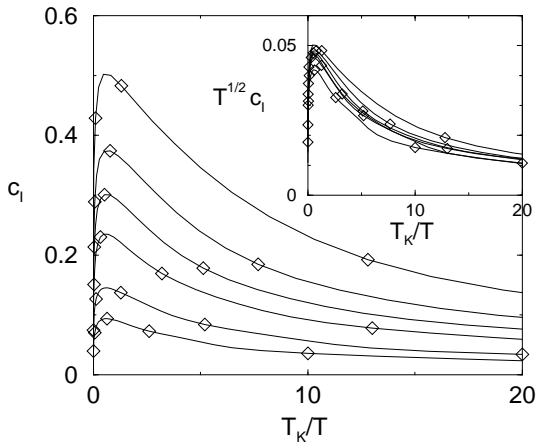}}
\caption{The coefficient $c_I$ for the two link impurity in
Eq.~(\ref{twolinkalt}) as a function of $T_K/T$ for different
temperatures $T/J = 0.2,0.1,0.04,0.025,0.0167,0.01$ from below.}
\label{cIscale}
\end{figure}}

\subsection{Doping in spin chains}
Our results also have immediate experimental consequences for impurities in 
spin-chain compounds such as KCuF$_3$ or $\rm Sr_2CuO_3$.
The spin density oscillations are directly linked
to the local Knight shifts (susceptibilities) close to the
corresponding impurities, which can be measured by standard NMR techniques
or muon spin resonance.  NMR experiments have already successfully
detected the sharp feature corresponding to the maximum 
in Fig.~\ref{altsusc} from open boundaries due to 
non-magnetic defects that were naturally present in the crystal.\cite{NMR}
We now propose to use intentional doping with magnetic or non-magnetic 
impurities to see the predicted
renormalization effects.  Impurities of one or
two modified links in the chain
can possibly be created by doping the surrounding 
non-magnetic atoms in the crystal at link or site parity symmetric 
locations.  The spin-1 impurities in Sec.~\ref{spin1} could be produced 
in a more straightforward way by 
substituting Cu ions by Ni ions in the corresponding compounds.
In Sec.~\ref{spin1} we discussed explicitly how the renormalization
effects for spin-1 impurities would show up in an actual experiment.
Similar arguments can also be made for the two link\cite{2CK}
or one link impurities by simply using the analytic form of 
the corresponding alternating spin densities with the 
coefficients $c_R$ and $c_I$ that we have calculated.

In general we find a strong enhancement of the antiferromagnetic order 
near impurities.  This enhancement can also be observed 
in higher dimensions\cite{elbio} and may have important consequences for
impurity-impurity interactions. In one dimension this effect is 
strongest, but the complex functional dependence we found here
is often beyond the intuitive explanation in terms of valence bond 
states.\cite{elbio}

\subsection{Impurities in Mesoscopic systems}
Finally, our analysis also allows us to draw important conclusions
for transport measurements in one dimensional mesoscopic structures.
This is probably the first time that the conductivity could be 
explicitly extracted from numerical data for Luttinger Liquid type models.
Not surprisingly, we found that a generic
impurity indeed renormalizes to complete backscattering as the temperature is 
lowered, and we also could explicitly observe the ``healing effect'' in  
the symmetric resonant tunneling case as predicted by Kane and 
Fisher.\cite{kane1,kanereview,kane2} Our numerical results
not only confirm the asymptotic power-laws, but also 
give a quantitative estimate of the conductivity for all temperatures
and impurity strengths.  For a generic impurity with little or 
intermediate backscattering we find that the asymptotic scaling region
turns out to be extremely narrow. For impurities with strong backscattering 
we find that the conductivity is {\it enhanced} by interactions at 
higher temperatures.

One obvious question is how those results can be generalized to spinful 
electron systems and carbon nanotubes.  A number of works have addressed
the question of impurities in spinful wires\cite{kanereview,yue,furusaki,wong} 
and found a richer structure since renormalization takes place in both the
spin and the charge channels.  However, if realistic SU(2) invariant 
interactions are assumed the generic behavior is very similar to the spinless
case, so that we expect that our results for the reflection coefficient 
carry over in a straight forward fashion.  The shape and amplitude
of the density oscillations, however, will in general be very different
for spinful electron systems.  For carbon nanotubes it has been shown that 
the Friedel oscillations impose a characteristic pattern that can be 
observed with scanning tunneling microscopy.\cite{mele}  For spinful wires
it is expected that the Friedel oscillations from an open end
can reveal the nature of the spin-charge separation in real space.\cite{stm}
Although our results do not allow for quantitative predictions of the 
density oscillations in spinful systems, 
we generally expect that strong, long-range density oscillations should
be present from backscattering in one-dimension.
One experimental consequence of those oscillations
is that the measurement through a lead close
to an impurity is very sensitive to the exact location.  
Previous studies have shown that even 
the distance between two leads can play a crucial role.\cite{kinaret}
The current
may be strongly enhanced or depleted, depending on if the distance 
to the impurity is a multiple of $2 k_F x$ or not.
Especially interesting are therefore experiments with an adjustable 
lead such as a tunneling tip.
The direct observation of those oscillations could give detailed 
information about both the nature of the impurity and also about the
interactions in the system.

\begin{acknowledgements}
S.R.\ acknowledges support from the Swedish Foundation for International
Cooperation in Research and Higher Education (STINT).  S.E.\ is thankful
for the support from the Swedish Natural Science Research Council 
through the research grants F-AA/FU 12288-301 and S-AA/FO 12288-302.
\end{acknowledgements}


\end{document}